\documentclass{article}
\usepackage{iclr2025_conference,times}

\usepackage{amsmath,amsfonts,bm}

\def\eqref#1{equation~\ref{#1}}

\def\1{\bm{1}}

\DeclareMathAlphabet{\mathsfit}{\encodingdefault}{\sfdefault}{m}{sl}
\SetMathAlphabet{\mathsfit}{bold}{\encodingdefault}{\sfdefault}{bx}{n}

\usepackage{hyperref}
\usepackage{url}

\usepackage{makecell}
\usepackage{array}
\usepackage{multirow}
\usepackage{color}
\usepackage{xcolor}
\usepackage{soul}
\usepackage[normalem]{ulem}
\definecolor{customred}{rgb}{0.98, 0.627, 0.627}
\definecolor{customgreen}{rgb}{0.816, 0.941, 0.753}

\newcommand{\hlred}[1]{{\sethlcolor{customred}\hl{#1}}}
\newcommand{\hlgreen}[1]{{\sethlcolor{customgreen}\hl{#1}}}

\title{Bridging Language Models and Financial Analysis}

\author{Alejandro Lopez-Lira \\
University of Florida \\
\texttt{Alejandro.Lopez-Lira@warrington.ufl.edu} \\
\And
Jihoon Kwon\\
LinqAlpha \\
\texttt{jihoonkwon@linqalpha.com} \\
\And
Sangwoon Yoon \\
Ministry of Justice, Republic of Korea \\
\texttt{asd01075272750@gmail.com} \\
\And
Jy-yong Sohn \\
Yonsei University \\
\texttt{jysohn1108@yonsei.ac.kr} \\
\And
Chanyeol Choi \\
LinqAlpha \\
\texttt{jacobchoi@linqalpha.com}
}

\begin{document}
\iclrfinalcopy

\maketitle
\pagestyle{plain}

\begin{abstract}
The rapid advancements in Large Language Models (LLMs) have unlocked transformative possibilities in natural language processing, particularly within the financial sector.
Financial data is often embedded in intricate relationships across textual content, numerical tables, and visual charts, posing challenges that traditional methods struggle to address effectively.
However, the emergence of LLMs offers new pathways for processing and analyzing this multifaceted data with increased efficiency and insight. Despite the fast pace of innovation in LLM research, there remains a significant gap in their practical adoption within the finance industry, where cautious integration and long-term validation are prioritized.
This disparity has led to a slower implementation of emerging LLM techniques, despite their immense potential in financial applications.
As a result, many of the latest advancements in LLM technology remain underexplored or not fully utilized in this domain.
This survey seeks to bridge this gap by providing a comprehensive overview of recent developments in LLM research and examining their applicability to the financial sector.
Building on previous survey literature, we highlight several novel LLM methodologies, exploring their distinctive capabilities and their potential relevance to financial data analysis.
By synthesizing insights from a broad range of studies, this paper aims to serve as a valuable resource for researchers and practitioners, offering direction on promising research avenues and outlining future opportunities for advancing LLM applications in finance.
\end{abstract}

\section{Introduction}

Recent advancements in Large Language Models (LLMs) have demonstrated significant potential in processing complex textual data with remarkable accuracy and scalability.
These models excel at handling vast amounts of information from diverse sources, understanding nuanced contexts in lengthy texts, and performing complex reasoning tasks that were once considered too difficult for machines.
This progress is especially relevant in the finance domain, where tasks often demand not only specialized domain knowledge but also sophisticated analysis of intricate data, making these challenges particularly complex.
In response, several researchers have begun exploring the use of LLMs to tackle these difficult financial tasks more effectively.

Although the potential of LLMs in finance continues to garner significant interest, an interdisciplinary gap continues to exist: finance and economics predominantly emphasize statistical inference and causal relationships, while computer science prioritizes predictive accuracy and scalability~\citep{lommers2021confronting}.
To be specific, Finance and economics often focus on understanding the underlying mechanisms behind observed phenomena, seeking to establish causal links through structured models and hypothesis-driven analysis~\citep{kumar2023causal, garg2024causal}.
On the other hand, computer science prioritizes optimizing algorithms for predictive accuracy and efficiency, frequently relying on large-scale data patterns sometimes without requiring explicit causal interpretation~\citep{domingos2012few}.
Establishing a collaborative environment that integrates technological expertise with financial domain knowledge is crucial to ensure that LLM-driven solutions remain accurate and aligned with real-world financial objectives.
This integration, through joint research, can enable robust, interpretable predictive models and foster innovation at the intersection of LLMs and finance.

In this survey, we aim to contribute to ongoing research by helping to address the gap between the rapid development of LLM technologies and their slower adoption in finance.
For this purpose, we highlight promising directions for future advancements and encourage further exploration at the intersection of LLMs and finance. 
Before exploring potential future directions, we first review the current literature on the use of LLMs in the finance domain, focusing on two key aspects: datasets (see ~\textbf{Section ~\ref{section_dataset}}) and models (see ~\textbf{Section ~\ref{section_models_application}}).
We then outline potential future directions by proposing how emerging LLM techniques could be applied to various areas of finance, as discussed in ~\textbf{Section ~\ref{section_future_direction}}.

Compared to existing survey papers that cover the use of LLMs in finance ~\citep{nie2024survey, lee2024survey, li2023large, dong2023scoping, zhao2024revolutionizing}, the main difference in this work lies in its introduction of the latest research on LLMs and its in-depth exploration of possible application areas within the finance domain.
While earlier surveys focused on well-established techniques that are already applied in finance, this survey offers both the technical update and guides on future research and innovation in financial LLMs.

\section{Textual Data in Finance Domain}
\label{section_dataset}
Financial data comes in various forms, including text, numerical, and time-series data, providing crucial insights for making informed financial decisions.
Despite the wealth of valuable information, textual data has historically been challenging to analyze using traditional methods that were developed before the advent of LLMs.
This difficulty arises because textual data, often collected from diverse sources such as company reports, news articles, and social media, requires strong interpretive language understanding with domain-specific knowledge.
Unlike traditional machine learning models, which have limitations in handling complex, unstructured text data, recent advancements in language models, particularly with LLMs, have demonstrated remarkable success in addressing these complexities more effectively~\citep{zhao2023survey, chang2024survey, minaee2024large, wei2022emergent, ogundare2023comparative}.
Here, we introduce four tasks where language models have been applied to financial textual data: text classification, information extraction, text summarization, and question answering.

\subsection{Text Classification}
In finance, text classification involves categorizing financial texts, such as news articles, earnings reports, and social media posts, into predefined categories based on content~\citep{tang2023finentity, alhamzeh2022s, jorgensen2023multifin}.
Two primary tasks in text classification are sentiment analysis and financial prediction.
Sentiment analysis classifies the sentiment conveyed in financial texts—typically as positive, neutral, or negative—helping to assess market sentiment or public opinion~\citep{malo2014good, shah2023trillion, lee2023stockemotions, casanueva2020efficient, chen2024efsa}.
Financial prediction focuses on forecasting outcomes such as market performance (e.g., stock price movements, market volatility, or risk)~\citep{sinha2021impact, xu2018stock, zhou2021trade, mathur2022monopoly, gerz2021multilingual, pei2022tweetfinsent, saqur2024nifty, soun2022accurate, wu2018hybrid} and corporate events~\citep{arno2024numbers, liang2020f, jacobs2018economic}, often framed as multi-labeled classification tasks.
Common evaluation metrics for financial text classification include accuracy and F1-score~\citep{baeza1999modern, van1979information}.

\subsection{Information Extraction}
Information Extraction (IE) in finance involves extracting structured information from unstructured financial texts. This process is crucial for analyzing company performance and tracking financial events. Named Entity Recognition (NER) and Relation Extraction (RE) have been important research topics in finance.
NER identifies specific entities in financial documents, such as companies, currencies, key financials, or dates~\citep{alvarado2015domain, loukas2022finer, sharma2023financial}. This task enables a better understanding of key components within financial texts by labeling this information. 
RE seeks to uncover the connections between these identified entities, such as relationships between financial figures and their respective events~\citep{sharma2022finred, kaur2023refind, mariko2020financial}. This task helps to extract actionable insights from texts by linking entities in meaningful ways.
Together, these tasks enhance the ability to process and analyze complex financial documents.
Evaluation metrics for Information Extraction tasks often include precision, recall, and F1-score~\citep{cowie1996information, sarawagi2008information}.

\subsection{Text Summarization}
Text Summarization in finance focuses on condensing lengthy financial documents, such as earnings reports, regulatory filings, and market analyses, into shorter, digestible summaries without losing key information~\citep{mukherjee2022ectsum, el2019multiling, zmandar2021financial, el2020financial, zavitsanos2023financial, liu2023long}.
This task is critical for helping analysts and investors quickly grasp essential insights from vast financial data.
Common evaluation metrics for text summarization include ROUGE~\citep{lin2004rouge} and BLEU~\citep{papineni2002bleu} scores, which measure the quality and accuracy of the generated summaries.

\subsection{Question Answering}
Question Answering (QA) in finance is designed to respond to complex questions leveraging financial knowledge and quantitative analysis.
financial QA often requires a deep understanding of relationships between different concepts, and the application of appropriate calculation methods~\citep{islam2023financebench, xu2024fintruthqa, krumdick2024bizbench, lai2024sec, reddy2024docfinqa}.
Recently, tasks combined with tabular data and unstructured text have gained traction~\citep{chen2021finqa, watson2024hiddentables, chen2022convfinqa, zhu2021tat, chen2024fintextqa, zhao2022multihiertt, liu2023tab, deng2022pacific}. These tasks require models to process and synthesize information from diverse sources to answer the question.
Common evaluation metrics for financial QA include Exact Match (EM)~\citep{rajpurkar2016squad, chen2019evaluating} and correctness~\citep{adlakha2023evaluating}, which measure the alignment between a model’s predicted answer and the ground truth, evaluating the accuracy of the response.

\section{Language Models for Diverse Financial Applications}
\label{section_models_application}
Language models are probabilistic models that assign probabilities to text sequences based on patterns learned from training data~\citep{hiemstra2009language}.
After the early efforts in language modeling that focused on rule-based approaches~\citep{weizenbaum1966eliza, bar1960present}, statistical methods such as N-gram and Hidden Markov models became prominent, which relied on word frequency and sequence patterns~\citep{brown1992class, fine1998hierarchical, markov2006example, eisenstein2019introduction}.
After the introduction of neural networks, deep learning-based models have achieved remarkable performance in language modeling~\citep{schuster1997bidirectional, graves2012long}, greatly enhancing the ability to understand and generate complex text sequences~\citep{li2018deep, li2022survey, o2020deep, bengio2000neural, graves2013generating, wang2015learning, sutskever2014sequence, sundermeyer2012lstm, bahdanau2014neural}.
Among these, models with Transformer~\citep{vaswani2017attention} architecture have been the most widely researched and utilized in recent years, demonstrating exceptional performance in tasks related to both language understanding~\citep{clark2020electra, devlin2018bert, joshi2020spanbert, he2020deberta} and generation~\citep{radford2018improving, radford2019language, brown2020language}.

Building on early work with the Transformer’s encoder architecture~\citep{devlin2018bert, clark2020electra, liu2019roberta} which excels in language understanding, several approaches have adapted this structure and its training methods for applications in the finance domain~\citep{araci2019finbert, liu2021finbert, yang2020finbert, huang2023finbert, shah2022flue, banerjee2024fine}.
With the emergence of LLMs utilizing decoder architecture~\citep{radford2018improving, radford2019language, brown2020language}, which have demonstrated remarkable performance in both language understanding and generation, these models have been increasingly applied to complex generation tasks in finance~\citep{wu2023bloomberggpt, yang2023fingpt, zhang2023instruct, liu2023fingpt, xie2023pixiu, yang2023investlm, li2023cfgpt, chen2023disc, zhang2023xuanyuan, bhatia2024fintral}.
These finance-specific models have shown superior performance in tasks involving financial text data, and their success has accelerated research into applying these technologies to the financial domain~\citep{zhu2023soargraph, banerjee2024fine, wang2023docgraphlm, do2024enhancing, feng2023empowering, xu2023instruction, wang2023finvis, mo2024fine, cao2024catmemo, su2024numllm, zhou2024silversight, inserte2024large, wang2023docllm, zhang2024finsql, chu2023data, son2023beyond}.
Recently, by altering the training objective to generate extended reasoning sequences, LLMs have demonstrated marked improvements in reasoning performance~\citep{gandhi2024stream, lightman2023let, snell2024scaling, jaech2024openai}, thereby expanding their potential in diverse financial applications.

In this section, we explore how language models have been applied across various financial areas.

\subsection{Trading}
Trading in finance encompasses all activities related to buying and selling assets or financial instruments.
There are two main types of investment analysis techniques in the stock market: technical analysis~\citep{achelis2001technical, edwards2018technical} and fundamental analysis~\citep{abarbanell1997fundamental, dechow2001short}.
While technical analysis predicts future price movements based on stock price trends, fundamental analysis evaluates a company’s intrinsic value based on its financial health and market outlook.
There have been various attempts to use machine learning for technical analysis, either to predict stock prices by examining the sentiment of public mood or news~\citep{bollen2011twitter, schumaker2009textual, tetlock2007giving} or by analyzing time series data~\citep{bao2017deep,sezer2020financial,torres2021deep}.

The development of LLMs has enabled further sophisticated \textit{sentiment analysis} for predicting stock prices by incorporating diverse financial document sources~\citep{lopez2023can, wu2024portfolio, kirtac2024sentiment, zhang2023unveiling, chen}.
These studies leverage the linguistic capabilities of LLMs to process unstructured data to improve the explainability and capture enhanced insights.
A common approach involves using LLMs to summarize or refine raw data, improving its quality and relevance for analysis.
LLMfactor~\citep{wang2024llmfactor} combined Chain-of-Thought (CoT) and Retrieval-Augmented Generation (RAG) for stock movement prediction.
MarketSenseAI~\citep{fatouros2024can} utilized external APIs with CoT and In-context Learning leveraging the reasoning capability of LLMs.
Some models also utilize summarized historical data as memory to manage complex financial data more effectively and deliver context-aware predictions~\citep{zhang2024multimodal, yu2024finmem}.
Further research includes employing multi-agent systems where LLMs debate predictions to enhance forecasting accuracy~\citep{li2023tradinggptmultiagentlayeredmemory, xing2024designing}.
Additionally, some researchers tried to combine reinforcement learning with LLMs to optimize performance, further advancing their predictive capabilities~\citep{explangpt, ding2023integrating}.
These developments highlight the transformative role of LLMs in financial sentiment analysis, enabling models to extract deeper insights and deliver more precise predictions.

LLMs have also been utilized for \textit{time series analysis}.
Recent studies highlight their ability to integrate textual and numerical data in stock forecasting and portfolio management.
For instance, StockGPT~\citep{mai2024stockgpt} prompts nearly a century of market data to predict stock returns by uncovering hidden patterns.
StockTime~\citep{wang2024stocktime} integrates textual and time series data to forecast stock prices across arbitrary look-back periods.
BreakGPT~\citep{zhang2024breakgpt} suggests a multi-stage approach to detect financial breakout by analyzing the stock time-series with LLMs.
FinVis-GPT~\citep{wang2023finvisgptmultimodallargelanguage} combines chart images and user queries to provide stock predictions with explanatory insights.
Alpha-GPT~\citep{wang2308alpha} and QuantAlpha~\citep{wang2024gholygrail} suggest systems that leverage human-AI collaboration and prompt engineering with large language models to identify market patterns capable of predicting market movements.
Beyond forecasting, LLMs have also advanced portfolio construction, surpassing traditional reinforcement learning models by tailoring advice to investor-specific contexts~\citep{etf} in contrast to primary method relied on reinforcement learning~\citep{etfrl}.
Additionally, LLMs used in~\citep{dimitrios} extract company embeddings from SEC filings, aligning with GICS classifications and correlating with financial metrics, enhancing financial analyses.

\textit{Fundamental analysis} in financial markets has historically been challenging for traditional language models~\citep{alexmaxvala}. Before the advent of Large Language Models (LLMs), several limitations hindered the application of language models in this domain. Traditional models struggled with processing and reasoning through extensive financial information, primarily due to their constrained ability to handle large contexts. These models also lacked the capability for \textit{deductive reasoning}, which is essential for interpreting complex financial data and market dynamics. Furthermore, the computational constraints of earlier models restricted the volume of input information, making comprehensive financial analysis impractical.
The emergence of LLMs has revolutionized fundamental analysis by addressing these limitations. Modern LLMs can incorporate extensive background knowledge through prompting techniques, while Chain of Thought (CoT) approaches enable sophisticated deductive reasoning capabilities~\citep{wei2023chainofthoughtpromptingelicitsreasoning}. The enhanced context processing abilities of LLMs have expanded the scope of fundamental analysis~\citep{alexmaxvala}, allowing for the simulation of diverse analyst perspectives~\citep{li2023tradinggptmultiagentlayeredmemory}. The integration of Retrieval-Augmented Generation (RAG) pipelines has further improved this capability by efficiently retrieving relevant documents based on specific inquiries~\citep{lewis2021retrievalaugmentedgenerationknowledgeintensivenlp}.
Recent research demonstrates the practical applications of LLMs in fundamental analysis. Gabaix et al.~\citep{gabaix} showcase how LLMs can effectively categorize firms and investors using their distinctive characteristics. Their approach employs vector representations, or asset embeddings, to capture firm- and investor-level features, demonstrating superior performance compared to conventional metrics in firm valuation, return co-movement analysis, and asset substitution pattern identification. These advancements have enabled the integration of qualitative narratives with quantitative measures, bridging a crucial gap in financial analysis.
The practical applications of LLMs extend further into analytical support tools. Research has shown that GPT's log probabilities can effectively quantify business complexity~\citep{bernard}. LLMs have proven valuable in simulating managerial responses during earnings calls~\citep{bae}, while ChatGPT has demonstrated effectiveness in evaluating corporate policies~\citep{jha}. These developments highlight the transformative impact of LLMs on fundamental analysis, offering new possibilities for comprehensive financial assessment and decision-making support.

\subsection{Financial Risk Modeling}
Financial risk modeling is crucial for predicting and mitigating potential financial losses, ensuring the stability of financial systems, and aiding in strategic decision-making. Effective risk modeling allows stakeholders to anticipate financial distress, fraud, and systemic risks, which are essential for maintaining investor confidence and regulatory compliance.

However, financial risk modeling is challenging due to the complexity and variability of financial data. Financial statements are vast and often contain unstructured, noisy data that require sophisticated methods to extract meaningful insights. Moreover, the dynamic nature of financial markets and the interplay of various economic factors add to the complexity. To address these challenges, various types of data must be analyzed, including corporate financial statements, disclosure texts, earnings call transcripts, and market-related time series data. The advent of language models has facilitated the analysis of large volumes of such financial text.

For example, Kim and Yoon~\citep{kimyoon} employ a BERT-based model to analyze the tone of corporate disclosure data (10-K) and predict corporate bankruptcy by integrating the results with additional financial variables. Similarly, Bhattacharya and Mickovic~\citep{fraud} utilize a BERT-based model to analyze 10-K reports for detecting accounting fraud. Recent research has further explored the application of LLMs in analyzing and measuring financial risks, including credit scoring~\citep{gencredit, sanz2024credit}, as demonstrated by Cao et al.~\citep{cao2024catmemo}.
Additionally, Lyu et al.\citep{systemicrisk} leverage OpenAI’s embedding model to construct a knowledge graph representing U.S. bank networks, enabling systemic risk assessment. Kim et al.~\citep{alexnik3} utilize GPT-4 to analyze earnings call data and integrate it with other economic variables to develop firm-level measures of political, climate, and AI-related risks. The advent of LLMs facilitates the integration of diverse data inputs, such as text, voice, and graphs, enabling more accurate results through comprehensive analysis.
For instance, Cao et al.~\citep{cao2024risklabspredictingfinancialrisk} predict financial risks by combining voice data from earnings calls, market-related time series data, and MD\&A text data. LLMs also enable multidimensional analysis by addressing new subtasks that are challenging to pre-train for. Similarly, RC2R~\citep{yu2024fusing}, which integrates LLMs with financial knowledge graphs, has been proposed to analyze causal mechanisms behind financial risk contagion. Finally, Choi and Kim~\citep{choikim} analyze narrative disclosures related to tax audits to examine their impact on corporate tax avoidance, investment decisions, and short-term stock prices.

\subsection{ESG Scoring}

ESG Scoring is a way to evaluate and quantify a specific company's performance and risks related to Environmental, Social, and Governance (ESG) factors. ESG scores are used by investors, companies, and policymakers to understand and compare a company's sustainability and social responsibility.

Prior to the advent of LLMs, various researchers have tried to utilize BERT-based models for ESG scoring. For example, BERT-based models are used to detect ESG-related issues in texts~\citep{pontes-etal-2023-leveraging}, analyze news data to assign ESG-related scores~\citep{haeinesg}, and predict the ESG impact duration and classify the types of influence of ESG-related factors~\citep{kim-etal-2024-leveraging-semi, banerjee-etal-2024-fine-tuning}. These efforts established a solid foundation for applying natural language processing in ESG tasks, showcasing the potential of structured approaches to analyze unstructured ESG data effectively.

With the advent of Large Language Models (LLMs), the field has witnessed significant advancements in ESG-related applications. LLMs have been utilized to predict ESG impact and event duration through in-context learning with GPT-4, where relevant examples are retrieved for each test instance, and instruction-tuning smaller models like Mistral (7B) with rationale generation, showcasing their effectiveness in ESG evaluations~\citep{rajpoot2024adapting}.
These capabilities have been further extended through multimodal approaches that integrate textual and structured data, enabling comprehensive ESG risk score prediction~\citep{nandiraju2024leveraging}.

Notably, LLMs have proven effective in extracting structured insights from sustainability reports, uncovering valuable regional and sector-specific patterns in corporate ESG disclosures~\citep{bronzini2024glitter}. The technology has also shown promise in multilingual applications, with generative LLMs demonstrating strong performance in classifying ESG issues across diverse language contexts~\citep{lee2023easyguide}.

Despite these advances, research on applying LLMs to ESG tasks remains in its early stages. Key challenges persist, particularly in ensuring scoring consistency and effectively addressing high-stakes scenarios. Further development in these areas will be essential for realizing the full potential of LLMs in ESG analysis.

\subsection{Mergers and Acquisitions (M\&A) Forecasting}
The application of language models has emerged as a valuable tool in predicting corporate mergers and acquisitions (M\&A). Early research demonstrates promising results using BERT-based architectures. For instance, Hajek and Henriques~\citep{manda} developed a model that analyzes the relationship between news sentiment and M\&A success rates, providing valuable insights into market dynamics and transaction outcomes.
Further advancing this field, Sawhney et al.~\citep{sawhney-etal-2021-multimodal} introduced a sophisticated approach combining a BERT-based language model with a customized transformer-based speech encoding model. This innovative methodology enables the extraction of financial risk indicators from M\&A conference calls, enhancing our understanding of transaction-related risks.
The emergence of Large Language Models (LLMs) has opened new possibilities for multimodal analysis in M\&A forecasting. While this represents a significant advancement over previous single-modal approaches, research leveraging these capabilities for M\&A outcome prediction remains an unexplored frontier, presenting opportunities for future investigation.

\subsection{Financial Agent}
The emergence of Large Language Models (LLMs) has revolutionized financial decision-making assistance through sophisticated agent-based systems. Research has demonstrated that the effectiveness of investment decisions heavily depends on the clarity and contextual integration of financial information~\citep{zang, alexnik2}. In response to this finding, modern financial agents are designed to deliver precise, contextually relevant information based on user queries.
Significant advances have been made in this domain. Kim et al.~\citep{alexmaxvalb} demonstrated that LLM-generated summaries effectively eliminate "disclosure bloat," providing investors with more concise and actionable information compared to original texts. Building on this foundation, Yang et al.~\citep{yang2024finrobotopensourceaiagent} developed FinRobot, a comprehensive toolkit for creating specialized financial LLM agents. Further expanding the field, Zeng et al.~\citep{zeng2024flowmindautomaticworkflowgeneration} introduced FlowMind, an innovative framework that generates workflows for financial tasks using natural language input, alongside NCEN-QA, a specialized dataset for N-CEN report analysis.
While NCEN-QA's primary focus extends beyond direct investment applications, it exemplifies the broader potential of LLMs in processing complex financial documents and supporting diverse financial decision-making processes.

\subsection{Lab Experiment Using LLM Agents}
The emergence of Large Language Models (LLMs) has introduced novel opportunities in experimental economics, offering an alternative to traditional human subject experiments. While conventional experiments face challenges such as IRB approval requirements, time constraints, costs, and recruitment difficulties, LLM agents provide a flexible and cost-effective solution for conducting experiments at any time and location.

Horton~\citep{experimenthorton} pioneered the use of LLM-based experiments as preliminary pilot studies for human subject research. Their work successfully replicated seminal studies by ~\citep{charness2002understanding, kahneman1986fairness, samuelson1988status}, achieving comparable results using LLM agents. Expanding on this foundation, Phelps and Russel~\citep{phelps2024machinepsychologycooperationgpt} revealed that through sophisticated prompt engineering, LLM agents could be endowed with psychological characteristics fundamental to behavioral economics, including altruism, cooperation, competitiveness, and selfishness.

In the financial domain, notable advances have emerged. StockAgent~\citep{zhang2024aimeetsfinancestockagent} introduced a multi-agent AI system powered by LLMs, designed to simulate real-world trading behaviors and analyze external factor impacts, providing valuable insights into trading dynamics and investment strategies. To evaluate the economic reasoning capabilities of LLMs, EconNLI dataset~\citep{econli} was developed as a comprehensive testing framework.
While the application of LLM agents in finance remains in its early stages, recent research by~\citep{gao2024simulatingfinancialmarketlarge} has demonstrated promising results. Their study implemented LLM agents in financial market simulations, utilizing time-series data of order history, stock holdings, and current balance to conduct various financial scenarios. The results closely aligned with established findings from traditional economics research, validating the potential of this innovative approach.

\section{Possible Future Directions: Applying Recent LLM Techniques in Finance Domain}
\label{section_future_direction}
The rapid advancements in LLMs have unlocked new opportunities for financial applications, yet many cutting-edge techniques remain underutilized in this domain. This section explores promising future directions for integrating recent LLM advancements into financial reasoning, decision-making, and analysis. Specifically, we examine how prompting techniques, RAG, tool-augmented LLMs, multimodal capabilities, multi-agent simulations, and model blending can enhance financial applications. While initial research has demonstrated the potential of these methods, their full impact on complex financial tasks yet to be realized. By systematically applying these innovations, LLMs can achieve greater accuracy, adaptability, and efficiency in financial modeling and prediction.

\begin{table}[h]
    \centering
    \begin{tabular}{|m{8cm}|}
    \hline
    \textbf{Question:} What is the sentiment of the statement "JPMorgan lowers the target price for Mengniu Dairy from HK\$51.8 to HK\$49, with an 'overweight' rating"? \\
    \hline
    \textbf{LLM's Answer (without examples)}: The sentiment is \hlred{balanced}, reflecting both a slight target price reduction and an optimistic "overweight" rating. \\
    \hline
    \textbf{LLM's Answer (with examples)}: The sentiment is \hlgreen{negative}, as the reduction in target price outweighs the positive impact of the "overweight" rating. \\
    \hline
    \end{tabular}
    \caption{An example showing impact of \textit{in-context learning} on Sentiment Analysis. Without examples, models may produce an \hlred{incorrect result}, while models can provide a more \hlgreen{accurate and context-aware evaluation} if few examples are provided. The question and examples used in the \textit{in-context learning} setting come from ~\citep{chen2024efsa}. Examples are generated by GPT-4o~\citep{achiam2023gpt}}
    \label{table:llm_in_context_learning}
\end{table}

\subsection{Prompting Techniques for Financial Reasoning}
Prompting techniques refer to the strategic design of task-specific instructions, known as prompts, to guide the output of LLMs without requiring additional model retraining or fine-tuning~\citep{liu2023pre, qiao2022reasoning, sahoo2024systematic}. These techniques enable models to dynamically adapt to a wide range of tasks by framing instructions in a way that optimizes their understanding and response capabilities~\citep{radford2019language, brown2020language}. LLMs can perform a variety of tasks~\citep{radford2019language} when carefully crafted prompts are used to guide them toward new tasks, a method known as \textit{zero-shot prompting}. Additionally, they can generalize more effectively through \textit{in-context learning}~\citep{brown2020language}, where the model is provided with a few input-output examples to induce an understanding of the target task.

Moreover, LLMs can also improve their reasoning abilities without the need for additional training~\citep{kojima2022large, wei2023chainofthoughtpromptingelicitsreasoning}. One effective technique is \textit{Chain-of-Thought} (CoT) prompting, which guides models to break down complex problems into step-by-step reasoning processes, thereby enhancing their logical consistency and accuracy~\citep{wei2023chainofthoughtpromptingelicitsreasoning}. Follow-up research has expanded on this concept through various approaches, such as Self-Consistency~\citep{wang2022self}, Tree-of-Thoughts~\citep{treeofthoughtsyao}, Graph-of-Thoughts~\citep{graphofthoughts}, and other related techniques~\citep{hu2023chain, zhou2022large, zhou2023thread, zhao2023enhancing}, further improving the model's ability to handle multi-step reasoning tasks effectively.

Despite their significant advantages, prompting techniques are still not fully utilized in LLMs for financial applications. While some research has explored prompting approaches, similar to CoT or zero-shot prompting, to enable systematically interpreting financial statements by LLMs~\citep{shaffer2024scaling, krause2023proper}, there are many LLM prompting techniques that remain largely underutilized. For example, \textit{in-context learning} can enhance LLM performance in tasks like sentiment analysis by using a few-shot prompting approach. As shown in Table~\ref{table:llm_in_context_learning}, this technique allows the model to correct inaccurate assessments by incorporating example-driven guidance for more precise evaluations.

\subsection{RAG for Knowledge-based Inference}
RAG~\citep{lewis2021retrievalaugmentedgenerationknowledgeintensivenlp} enhances large language models (LLMs) by addressing key limitations such as hallucinations with efficient context windows~\citep{kang2023deficiencylargelanguagemodels}. RAG achieves this by retrieving external documents and grounding the model’s responses in verified, up-to-date knowledge, thereby improving the accuracy and relevance of its outputs without the need for extensive fine-tuning~\citep{jiang2023active, gao2023retrieval}. In the financial domain, where accurate and timely information is essential, RAG has shown promise its effectiveness but remains largely limited to few tasks, such as question-answering and sentiment analysis.~\citep{sarmah2023reducinghallucinationextractinginformation, zhang2023enhancingfinancialsentimentanalysis, iaroshev2024evaluating, zhao2024optimizing}. Thus, there are opportunities for further research to apply RAG to more complex financial tasks.

One particularly promising avenue for RAG in finance lies in tasks that require synthesizing multiple data sources to generate accurate and timely insights. Credit evaluation and risk management heavily rely on diverse and evolving information, including borrower transaction histories, macroeconomic indicators, alternative credit scoring data, and regulatory filings~\citep{bhattacharya2023credit, kedia2024credit, munoz2023dynamics, feng2023empowering}. Traditional models often struggle to incorporate and interpret such heterogeneous data in real time~\citep{kalapodas2006credit, faheem2021ai, bello2023machine}, but RAG can dynamically retrieve relevant external documents and integrate them with proprietary financial records to improve decision-making accuracy. Similarly, investment evaluation and portfolio management involve analyzing structured market data alongside unstructured information such as earnings reports~\citep{hunton2006financial}, news sentiment~\citep{malandri2018public, hung2024intelligent}, and geopolitical developments~\citep{papic2020geopolitical}. By leveraging RAG to process these disparate sources efficiently, financial analysts can enhance asset valuation, optimize portfolio allocation, and mitigate risks stemming from emerging market trends. As financial applications increasingly demand real-time, explainable, and context-aware insights, RAG presents a scalable and interpretable approach to addressing these challenges.

\subsection{Tool-augmented LLMs for Complex Financial Analysis}
Tool-augmented LLMs enhance the capabilities of language models by integrating external tools, enabling them to handle complex tasks and analyses. For more domain-specific adaptation, these models are optimized for specific tools or software, either through specialized training~\citep{qin2024toolllm, tang2023toolalpacageneralizedtoollearning} or by using external APIs or functions during inference~\citep{hsieh2023tooldocumentationenableszeroshot, paranjape2023artautomaticmultistepreasoning}. For example, an LLM can be tailored for a specific programming language, data analysis tool, or industry sector.

The importance of accessing specific tools and software in finance applications has grown. Research has introduced techniques to enhance external knowledge usage~\citep{parisi2022talmtoolaugmentedlanguage}, reasoning abilities~\citep{das2024mathsenseitoolaugmentedlargelanguage, gou2024tora}, and programming and arithmetic skills during inference without extensive pre-training~\citep{pal}. Additionally, studies explore how LLMs can use tools for real-time interaction, fostering ecosystems and platforms that increase tool usage~\citep{xu2024on, tang2023toolalpacageneralizedtoollearning, liang2023taskmatrixaicompletingtasksconnecting, xu2023gentopiacollaborativeplatformtoolaugmented}.

Applying this research to finance could enable complex financial analyses, such as valuation or financial modeling, by integrating real-time data, specific programming, or external APIs. However, research on using these tool-augmented LLMs for complex financial analyses remains limited. In finance, tool-augmented LLMs have primarily enhanced simple analytical tasks or utilized basic external APIs~\citep{theuma2024equippinglanguagemodelstool, zeng2024flowmindautomaticworkflowgeneration}. Therefore, we propose using tool-augmentation to enable LLMs to perform complex tasks requiring real-time data, specific programming, or external APIs.

\subsection{Multimodal LLMs for Understanding Non-Textual Data}
Multimodal LLMs, such as GPT-4~\citep{achiam2023gpt}, are increasingly used for extracting answers from documents that are not digitized or are difficult to process as text due to tables or images~\citep{yang2023dawnlmmspreliminaryexplorations}. In finance, critical materials for investment analysis, like annual reports and financial statements, often include tables, charts, and other visual elements. Although many studies in Chapter~\ref{section_models_application} utilized LLMs to analyze financial data, visual materials were often excluded due to the challenges of multimodal research.

Research on reading and comprehending layout-based materials and multimodal LLMs focused on visual data suggests potential applications in finance. Recent studies on document-level layout comprehension indicate that LLMs are increasingly capable of analyzing lengthy disclosure documents~\citep{xu-etal-2021-layoutlmv2, wang2023docllmlayoutawaregenerativelanguage, li2024enhancingvisualdocumentunderstanding}. Additionally, emerging research on visually understanding tables, systematically analyzing table data, and retrieving appropriate chart materials shows promise for finance applications~\citep{zheng2024multimodaltableunderstanding, chainoftable, tian2024spreadsheetllmencodingspreadsheetslarge, nowak-etal-2024-multimodal}. Ongoing research on chart comprehension and reasoning further enhances these capabilities~\citep{masry-etal-2023-unichart}.

While the application of multimodal research in finance is still in its early stages, it has the potential to extract financial insights from disclosure documents~\citep{wang2023finvisgptmultimodallargelanguage, bamford2024multimodalfinancialtimeseriesretrieval, li2023multimodal}. Recent studies employing diverse modalities hold promise for unveiling new insights within these documents.

\subsection{Multiagent-based Simulation}
Although various studies have used LLMs to model human behavior~\citep{wawer2024large, xiao2024tradingagents, yu2024fincon}, our understanding of how accurately these models replicate the decision-making processes of human economic actors remains limited~\citep{ross2024llm, sreedhar2024simulating}. Studies in game theory and behavioral economics have demonstrated that LLMs can exhibit behaviors resembling human economic reasoning~\citep{fontana2024nicer, kitadai2024can, chuang2024beyond}. In light of this potential, recent studies suggest employing multiple LLM agents as experimental tools to investigate their effectiveness in economic reasoning and behavioral simulations~\citep{albert2024reproducing, li2024econagent, experimenthorton}.

Understanding the behavioral patterns of LLMs is particularly crucial in finance, where modeling market participants requires a deep grasp of decision-making dynamics. Financial markets consist of diverse actors—such as investors, traders, and institutions—each driven by complex motivations and responses to market stimuli. If LLMs are to effectively simulate these participants, it is essential to evaluate whether their decision-making aligns with observed human behavior under varying economic conditions. Without this understanding, predictions of market reactions, risk assessments, and strategy optimizations could be misaligned, limiting the practical application of LLM-driven models in financial simulations and forecasting. Thus, further exploration of LLM behavior in finance-specific scenarios is needed to enhance their utility as reliable tools for modeling market interactions.

\subsection{Domain-Adaptation and Enhanced Efficiency through Model Blending}
While research predominantly focuses on LLMs, there is a growing need for domain-adapted small LMs. Even state-of-the-art LLMs have not outperformed BERT-based models in all tasks. Notably, research highlights that GPT-4 performs worse than RoBERTa-based models in many areas, including financial tasks~\citep{bertthangpt}. Since encoder-based small LMs and LLMs excel at different tasks, developing an integrated framework, or model blending, is essential. Research on blending between LLMs provides valuable insights~\citep{blending}.

In domain-specific areas like financial applications, LMs often require training on various task-specific datasets. Real-time demands in some financial sectors necessitate fast training and inference, where even small models like BERT may need GPU access for sub-second latency. Techniques like quantization and pruning make LMs more lightweight but can degrade performance as task complexity increases~\citep{quantization, pruning}. Thus, exploring optimal optimization levels for finance-related tasks and methods to minimize performance loss during model compression is crucial.

\section{Limitations of Using LLMs for Finance}
\label{section_limitations}

Despite the significant advancements and applications of LLMs in the financial domain, several critical limitations persist. These limitations must be addressed to enhance the reliability and effectiveness of LLMs in finance-related tasks. For example, as illustrated in Table~\ref{tab:failure_case_section_5}, hallucinated insights in equity research or financial statement can create confusion among investors by referencing non-existent company patents. Similarly, flawed mathematical reasoning during portfolio optimization can lead to a misrepresentation of risk and returns, potentially jeopardizing investment decisions.

\subsection{Hallucinations}
Financial systems utilizing LLMs are composed of numerous complex modules. Consequently, hallucinations—erroneous outputs generated by the model—can propagate through the system, causing significant failures. Hallucinations are categorized into factual and logical types~\citep{hallukind2}. Factual hallucinations are prevalent in tasks like Financial Abbreviation Recognition, Financial Term Explanations, and Stock Price Queries, appearing in at least 10\% of the samples for models such as GPT-4, Llama 2, and FinMA~\citep{hallufin}. Logical hallucinations, involving flawed deductive reasoning, receive less attention but are equally problematic~\citep{deduction}. These issues deter researchers from employing LLMs in experimental studies. Techniques like RAG can mitigate factual hallucinations but face limitations in specialized domains and when reasoning requires multiple document references. Prompting techniques such as Chain of Thought (CoT) and Tree of Thoughts (ToT) offer improvements in logical reasoning but are not comprehensive solutions, especially in experimental settings where ideal prompts cannot always be provided.

\begin{table*}[ht]
\centering
\caption{Potential Failure Cases in LLM-Driven Financial Analysis. LLMs may generate \hlred{erroneous outputs}, potentially leading to confusion and incorrect decisions.}
\label{tab:failure_case_section_5}
\begin{tabular}{p{5cm} p{8cm}}
\hline
\textbf{Type of Analysis and Error} & \textbf{Example of Erroneous Generation} \\
\hline

\textbf{Equity Research\newline(Hallucination)} 
& \textbf{Q}: Does Morgan Stanley hold any blockchain-related patents?\newline \textbf{A}: Yes, Morgan Stanley holds \hlred{U.S. Patent No. 10,987,654}, granted in \hlred{August 2023}, which focuses on \hlred{“multi-layer distributed ledger networks for high-frequency trading.”} According to the company’s Q2 2024 Investor Report, this technology allows for \hlred{“seamless cross-border asset settlement with real-time validation.”} \\
\hline
\textbf{Portfolio Optimization\newline(Mathematical Reasoning)} 
& \textbf{Q}: How do you calculate the variance of a two-asset portfolio with equal weight?\newline \textbf{A}:  Consider two assets, Asset A and Asset B, with variances of  0.04  and  0.09 , and a correlation of  0.5. then,

$\sigma_p^2 = 0.5^2 \times 0.04 + 0.5^2 \times 0.09 +$ \hlred{$0.5 \times 0.04 \times 0.09 = 0.0625$}\newline
Thus, the portfolio standard deviation is: $\sigma_p =$ \hlred{$\sqrt{0.0625} \approx 0.25$}
\\

\hline
\end{tabular}
\end{table*}

\subsection{Mathematical Reasoning}
The domain of finance relies heavily on mathematical reasoning, yet LLMs exhibit significant variability and inconsistency in this area. Even advanced models struggle with different textual presentations of mathematical problems~\citep{mathsurvey}. Self-consistency methods, which paraphrase problems into various forms, can improve performance but require reliable model performance~\citep{yoonkimmath}. This issue is exacerbated in languages with limited training resources. Furthermore, LLMs often reach correct answers through incorrect reasoning, undermining trust in their outputs. The CHAMP dataset highlights this issue, revealing that state-of-the-art LLMs frequently use flawed reasoning steps to arrive at correct answers~\citep{yoonkimmathdata}.

\subsection{Evaluation: Human in the Loop}
Evaluating LLM-generated outputs in finance relies heavily on human judgment due to the persistent issue of hallucinations. Although research into automated evaluation using LLMs is ongoing, these methods still depend on human oversight~\citep{geval, chateval}. The probabilistic nature of LLMs and frequent model updates introduce variability, reducing the robustness of evaluation results~\citep{gevalrepli}. Automated evaluation research has primarily focused on tasks with discrete scoring systems, leaving performance in continuous spaces unverified. This limitation underscores the need for robust human evaluation frameworks to ensure accuracy and reliability in LLM outputs.

\section{Conclusion}

This paper presents a comprehensive overview of language model evolution since the introduction of transformer architectures, with a particular focus on their applications in financial domains. We examine various case studies where language models have been successfully deployed to address financial challenges. While both computer science and finance researchers actively investigate financial Large Language Models (LLMs), their research objectives and methodological approaches exhibit distinct characteristics.

Finance researchers primarily approach language models as analytical tools for addressing fundamental questions within their domain. Their focus centers on the practical efficacy of these models in solving specific financial problems. In contrast, computer science researchers prioritize advancing the technical capabilities of language models, using financial applications primarily as demonstration cases for achieving state-of-the-art (SOTA) performance on established computer science benchmarks and downstream tasks.

The intersection of these research interests lies in addressing significant financial questions that remain unresolved due to current limitations in language model capabilities and scope. Successfully addressing these challenges would represent substantial progress in both social science research and technical engineering advancement. This paper identifies these critical questions and explores potential technical approaches for their resolution. Our survey aims to serve as a comprehensive resource for two distinct audiences: finance researchers seeking to integrate LLMs into their investigations, and computer science researchers pursuing applied research opportunities in finance.

\bibliography{main}
\bibliographystyle{iclr2025_conference}

\end{document}